\documentclass[aps,floatfix,twocolumn,showpacs,preprintnumbers]{revtex4-2}

\usepackage{graphicx} 
\usepackage{bm} 
\usepackage{amsmath,amssymb}
\usepackage{xcolor}

\graphicspath{ {./}{figs/}}
\begin{document}

\title{Absence of backscattering in Fermi--arc--mediated conductivity of
topological Dirac semimetal Cd$_{3}$As$_{2}$}
\author{Vsevolod Ivanov$^{1,2,3}$ Lotte Borkowski$^4$, Xiangang Wan$^5$,
Sergey Y. Savrasov$^4$}
\affiliation{$^1$Virginia Tech National Security Institute, Blacksburg, Virginia 24060,
USA}
\affiliation{$^2$Department of Physics, Virginia Tech, Blacksburg, Virginia 24061, USA}
\affiliation{$^3$Virginia Tech Center for Quantum Information Science and Engineering,
Blacksburg, Virginia 24061, USA}
\affiliation{$^4$Department of Physics, University of California, Davis, CA 95616, USA}
\affiliation{$^5$National Laboratory of Solid State Microstructures, School of Physics
and Collaborative Innovation Center of Advanced Microstructures, Nanjing
University, Nanjing, China. }

\begin{abstract}
Having previously been the subject of decades of semiconductor research,
cadmium arsenide has now reemerged as a topological material, realizing
ideal three-dimensional Dirac points at the Fermi level. These topological Dirac points lead to a number of
extraordinary transport phenomena, including strong quantum oscillations,
large magnetoresistance, ultrahigh mobilities, and Fermi velocities
exceeding graphene. The large mobilities persist even in thin films and
nanowires of cadmium arsenide, suggesting the involvement of topological
surface states. However, computational studies of the surface states in this
material are lacking, in part due to the large 80-atom unit cell. Here we
present the computed Fermi arc surface states of a cadmium arsenide thin
film, based on a tight-binding model derived directly from the electronic
structure. We show that despite the close proximity of the Dirac points, the
Fermi arcs are very long and straight, extending through nearly the entire
Brillouin zone. The shape and spin properties of the Fermi arcs suppress
both back- and side- scattering at the surface, which we show by explicit
integrals over the phase space. The introduction of a small
symmetry-breaking term, expected in a strong electric field, gaps the
electronic structure, creating a weak topological insulator phase that
exhibits similar transport properties. Crucially, the mechanisms suppressing
scattering in this material differ from those in other topological materials
such as Weyl semimetals and topological insulators, suggesting a new route
for engineering high-mobility devices based on Dirac semimetal surface states.
\end{abstract}

\maketitle

\section{Introduction.}


Cadmium arsenide (Cd$_{3}$As$_{2}$) is a well known semiconductor that has
been thoroughly studied for the greater part of a century for its remarkable
transport properties \cite{struct1935}. More recently, this material was
predicted \cite{Wang2013-kp} and confirmed using angle-resolved
photoemission spectroscopy \cite{Cava2013, Neupane2014-arpes, Yi2014-arpes,
Roth2018-arpes} to be a topological Dirac semimetal . A conventional Dirac
semimetal exists at the boundary between topological insulator (TI) and
normal insulator phases in a system possessing both time-reversal and
inversion symmetries, and hosts a Dirac node at a time-reversal invariant
momentum point in the Brillouin zone (BZ) \cite{Yan2017_review}. However,
such a phase can also exist when the crystal structure possesses a
rotational axis \cite{Gao2016}, such as the $C_{4}$ rotation in Cd$_{3}$As$%
_{2}$, which can result in accidental band crossings at the Fermi energy,
leading to two topologically protected Dirac points along the $\Gamma -Z$
direction \cite{Wang2013-kp}.

The properties of cadmium arsenide have been well established through a
number of experimental works, in which it served as a prototypical system
for demonstrating the optical conductivity \cite{PhysRevB.93.121202,
PhysRevB.94.085121, PhysRevLett.124.166404}, quantum oscillations \cite%
{PhysRevB.96.121407, Miyazaki2022, Zhang2017}, electronic transport \cite%
{Moll2016, cd3as2-fet, PhysRevX.5.031037, Liang2015, PhysRevLett.124.116802,
PhysRevB.104.205427, Baba_2021} of Dirac semimetal materials. Dirac
semimetals were additionally predicted to exhibit proximity-induced
superconductivity \cite{PhysRevLett.115.187001}, and Cd$_{3}$As$_{2}$ was
used in experimental realizations of this effect \cite{Wang2016,
PhysRevB.97.115446,Huang2019,PhysRevB.99.125305}. 

Cadmium arsenide also
received limited theoretical treatment using models \cite{Wang2013-kp,
Pan2015} and first-principles calculations \cite{Wang2013-kp,
MoscaConte2017, Mazhar2014, Baidya2020, cd3as2_wannier}, which are inherently limited by the
large number of atoms in its crystal unit cell. In particular, this makes
first-principles investigations of the surface state physics of Cd$_{3}$As$%
_{2}$ especially difficult on account of the even larger supercells
involved. 

The Dirac points in Cd$_{3}$As$_{2}$ can be considered as composed of
opposite chirality Weyl points, which are connected by Fermi arcs at the
material surface \cite{pyrochore-iridates}. It has been shown that the Fermi
arcs in Weyl semimetals can be highly conductive \cite{Resta2018}, prompting
the question of whether the Fermi arc states of Dirac semimetals can lead to
similar transport effects. This question is particularly relevant for the
case of cadmium arsenide, which has been shown to exhibit extraordinarily
high electronic transport even in thin-film and nanowire geometries \cite%
{cd3as2-fet, Nakazawa2018-struct, PhysRevB.96.121407,
PhysRevLett.124.116802, wang2022schottky}, where surface effects would be
expected to dominate.

Here we perform a numerical study of the electronic transport of Cd$_{3}$As$%
_{2}$ mediated by its Fermi arc surface states. We begin by empirically
deriving a tight--binding model fit to the first-principles electronic
structure, and use it to construct a 40-layer slab supercell to extract the
surface energy bands. We show that contrast to prior $k\cdot p$ models \cite%
{Wang2013-kp}, the Fermi arcs of Cd$_{3}$As$_{2}$ in our \textit{ab initio}%
-based model are long and straight, extending across the entire BZ to
connect across the zone boundary. By performing explicit calculations of the
phase space available to electronic scattering we demonstrate that two major
scattering contributions become vanishingly small regardless of the scattering
mechansim. This can potentially lead to very high conductance in  thin films of Cd$_{3}$As$_{2}$. We further extend this argument in the
presence of an external electric field, where we show that the induced
symmetry--breaking results in a weak topological insulator phase, whose
surface states preserve the essential properties necessary for high
electronic mobility. We conclude by discussing how these results might lead to next-generation electronic devices based on the high electronic mobility of the surface states and their potential hydrodynamic transport, and how such long, straight Fermi arcs might be engineered in semiconductor heterostructures.

\section{Material Background.}



Although research on the properties of Cd$_{3}$As$_{2}$ has been
reinvigorated by the confirmation of its topological electronic structure 
\cite{Cava2013}, the material itself has been known to exist for nearly a
century \cite{struct1935}. At high temperatures, cadmium arsenide adopts a
ten--atom anti--fluorite structure ($Fm\bar{3}m~\#225$) with two cadmium
vacancies \cite{struct1935}. Upon cooling to $\sim$ 600$^{\circ }$C, the
remaining cadmium atoms displace towards the ordered vacancies, resulting in
a the intermediate structure with a distorted $\sqrt{2}\times \sqrt{2}\times
2$ supercell ($P4_{2}/nmc~\#137$) of the original cubic structure \cite%
{struct1969}. Below $\sim$ 475$^{\circ }$C, the material settles into its
low--temperature phase, which is a $2\times 2\times 4$ superstructure of the
ten--atom anti-fluorite cell. The exact nature
of this low temperature phase has been somewhat controversial, initially
being assigned to the inversion-broken $I4_{1}cd$ ($\#110$) space group \cite%
{struct1968}, which was recently reexamined and found to actually be the
centrosymmetric $I4_{1}acd$ ($\#142$) \cite{Mazhar2014}. This has important
consequences for the electronic structure, preserving the Dirac crossings in
the BZ, and enabling the possibility of tuning the
topological phase by breaking inversion symmetry.

\begin{figure}[ht]
\includegraphics[height=1.02\columnwidth,width=1.0\columnwidth]{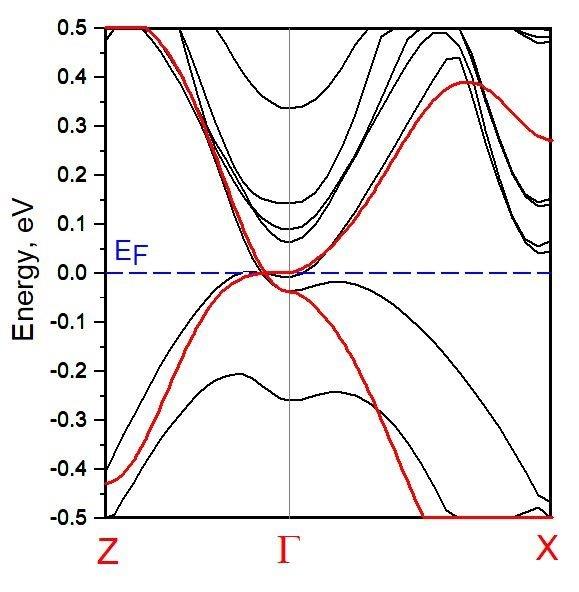}
\caption{Electronic structure of Cd$_{3}$As$_{2}$ derived from first
principles density functional based calculation(black), and its tight
binding fit (red).}
\label{bands}
\end{figure}

The topological nature of the electronic structure in Cd$_3$As$_2$ has been confirmed experimentally with transport and angle-resolved photoemission spectroscopy (ARPES) measurements. 

Electronic excitations around the Dirac cones at the Fermi level of Cd$_3$As$_2$ behave like relativistic Dirac fermions with linear dispersion. These excitations have extremely large Fermi velocities \cite{cd3as2-fet}, and are expected to lead to large electronic mobilities which have been observed experimentally \cite{Liang2015, PhysRevLett.124.116802, cd3as2-fet}. Surprisingly, these high electron mobilities persist even in thin--films and nanowires of Cd$_3$As$_2$ \cite{Pan2015, wang2022schottky, Nakazawa2018-struct, cd3as2-fet}. Bulk-boundary correspondence in topological materials \cite{Fu2007} implies that topological features in the bulk exist in tandem with topologically protected surface states, which in the case of bulk Dirac points take the form of Fermi arcs connecting their projections on the surface. Such Fermi arcs are known to be highly conductive for related Weyl semimetal systems \cite{Resta2018}, and would be expected to dominate the transport properties of Cd$_3$As$_2$ in thin--film and nanowire geometries.

ARPES measurements enable the direct observation of bulk and surface states. This has allowed the linearly dispersing Dirac cones at the Fermi surface to be directly observed  \cite{Cava2013, Neupane2014-arpes, Yi2014-arpes,
Roth2018-arpes}.  The derived $\bm{k}\cdot\bm{p}$ model \cite{Wang2013-kp} predicts the existence of short Fermi arcs enclosing the $\Gamma$ point, and later studies of the Weyl--semimetal state in symmetry--broken Cd$_3$As$_2$ using a model derived from Wannier interpolation also predict short arcs \cite{cd3as2_wannier}. However, the $\bm{k}\cdot\bm{p}$ model necessarily represents only the behavior near the $\Gamma$-point, poorly capturing the electronic structure at the BZ edge, and experimental ARPES measurements have yet to observe Fermi arc surface states. It has been suggested that the difficulty in observing Fermi arc surface states in Dirac semimetals might stem from their lack of topological protection \cite{Kargarian2016}, and in fact there exist Dirac systems that completely lack Fermi arcs \cite{Le2018}.

Directly simulating these surface states from first principles would be computationally prohibitive, as such a calculation would require a slab supercell consisting of at least several dozen enormous unit cells of Cd$_3$As$_2$. The inclusion of spin-orbit coupling in order to correctly capture the topological electronic structure, and the dense $\bm{k}$-point grids needed to obtain the Fermi surface would further complicate this approach. On the other hand, a $\bm{k}\cdot\bm{p}$ model, while quick to compute, would not be able accurately represent the electronic structure throughout the entire BZ. The most optimal approach, which admits a trade--off of speed and accuracy, is a tight-binding model.

\begin{figure}[ht]
\includegraphics[width=1.0\columnwidth]{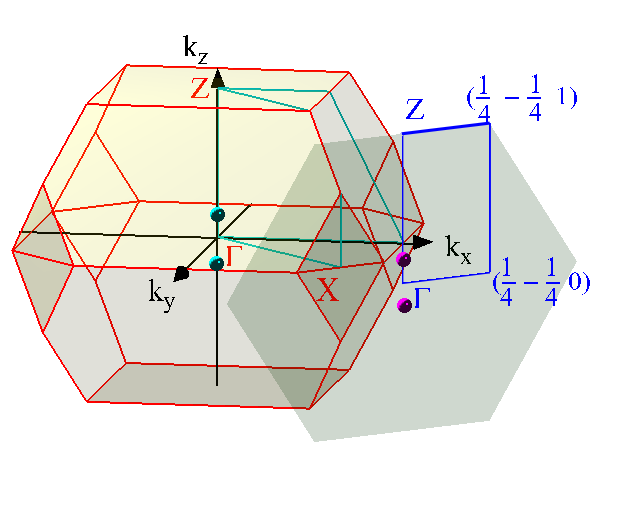}
\caption{The Brillouin Zone (BZ) of bulk Cd$_3$As$_2$ and its projection along the $langle 110\rangle$ direction that is used for calculating the surface states. The blue rectangle shows the portion of the surface BZ that will be used in subsequent plots. The bulk Dirac points along $\Gamma - Z$ and their projections onto the surface BZ are indicated by small circles.}
\label{bz}
\end{figure}

\section{Electronic Structure of C\lowercase{d}$_{3}$A\lowercase{s}$_{2}$}

We perform self-consistent density functional--based electronic structure
calculation of Cd$_{3}$As$_{2}$ using the full potential linear muffin--tin
orbital method (FPLMTO) including spin--orbit coupling \cite{Savrasov1992}.
The band structure plotted along the $Z-\Gamma -X$ direction
shown in Fig. \ref{bands}, clearly reveals the four-fold degenerate Dirac
point along the $\Gamma -Z$ direction which is protected by the $C_{4}$
rotational symmetry of the crystal \cite{Gao2016}, and is consistent with previous studies of Cd$_3$As$_2$ \cite{Mazhar2014}.

In order to numerically study the topological surface states of Cd$_{3}$As$%
_{2}$ and their transport properties, we derive a tight--binding model by
empirically fitting the electronic structure. Our model captures the essence
of the electronic behavior throughout the (BZ), and
accurately reproduces the four-fold topological Dirac point along the $%
\Gamma -Z$ direction as shown in red in Fig. \ref{bands}. (We describe the tight-binding fit and list the
parameters of the model in the Supplemental Material)

Using this tight--binding model we construct a 40-layer slab in the $\langle
110\rangle $ direction, by first extending the model over the supercell and
then forbidding hoppings between the top and bottom surfaces. The relationship between Brillouin Zones 
for the bulk and the slab is shown in Fig. \ref{bz}.

\begin{figure}[hb]
\includegraphics[width=1.0\columnwidth]{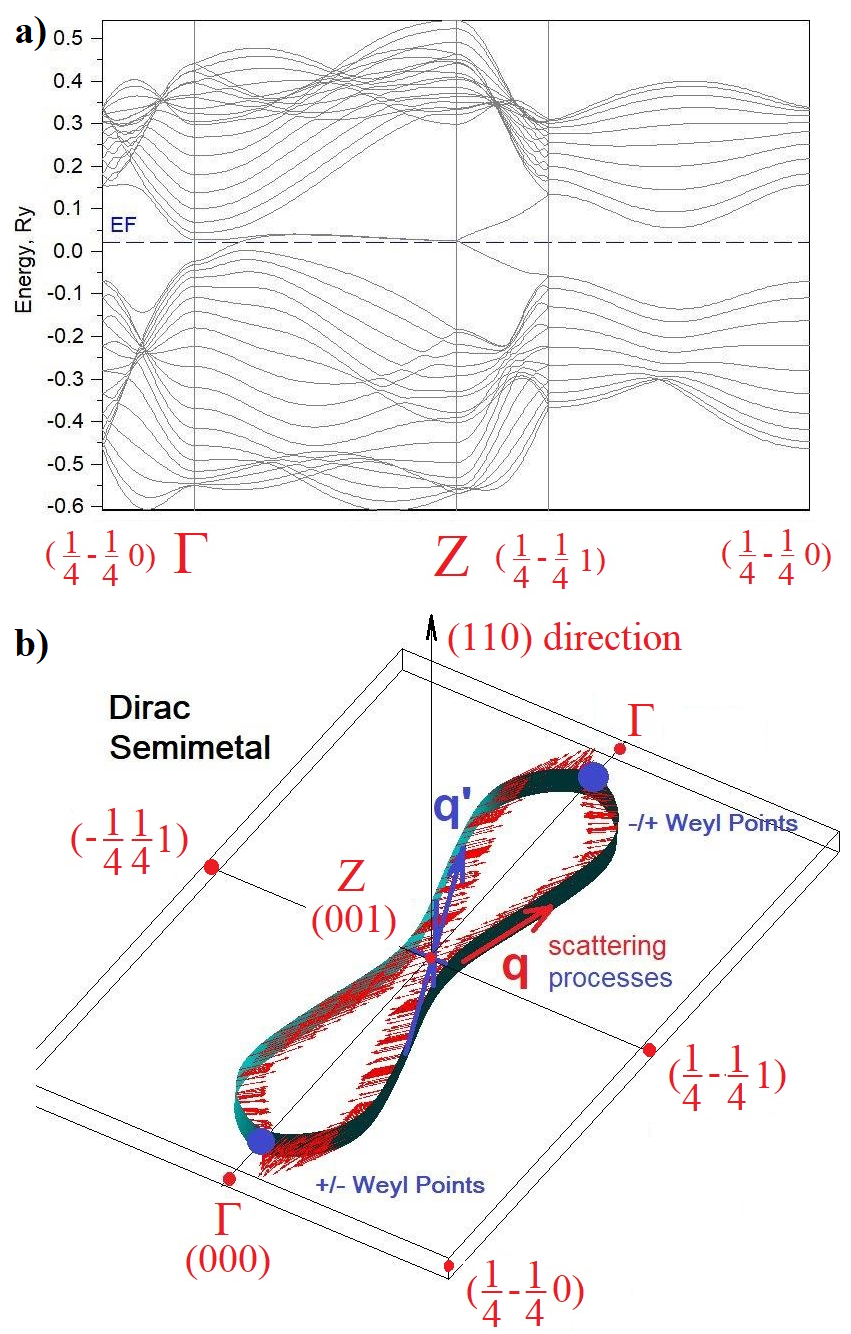}
\caption{40-layer slab calculation of the tight-binding model fit for the Dirac semimetal phase of Cd$_{3}$As$_{2}$, showing the a) band structure, and b) Fermi arc surface states. Note that the shown region is stretched along the $k_{\langle
1-10\rangle }$ direction in order to more clearly show the arc shape. }
\label{dirac-combined}
\end{figure}

The resulting
band structure can be plotted within the plane spanned by the $k_{\langle
1-10\rangle }=k_{xy}$ and $k_{\langle 001\rangle }=k_{z}$ vectors, shown in
Figure \ref{dirac-combined}a). The surface states arising from the topological Dirac
nodes are clearly visible as the only states crossing the Fermi energy.

Plotting the Fermi surface of the slab calculation reveals that the two
Dirac point projections are connected by very long, straight Fermi arcs
(Fig. \ref{dirac-combined}b). In contrast to $k\cdot p$ calculations \cite{Wang2013-kp}%
, these arcs do not enclose the $\Gamma $ point, and instead extend through
the entire BZ to connect across the BZ boundary. The origin of the arcs can
be understood in terms of the $\pm $ chirality Weyl points comprising each
Dirac point. Each pair of Weyl points located at opposite momentum space $k$%
-points are connected by a single Fermi arc. The two Dirac points of Cd$_{3}$As$_{2}$ can be thought of as two pairs of coincident Weyl points of
anti-aligned chiralities, with each pair having its own Fermi arc. An
important feature of the Fermi arcs is the spin-texture; spins along each
arc point along opposite directions, but rotate as they approach the Dirac
node projections to align with local \textquotedblleft
all-out\textquotedblright\ arrangement around each point.


\section{Transport Analysis}

\begin{figure*}[tb]
\includegraphics[width=1.0\textwidth]{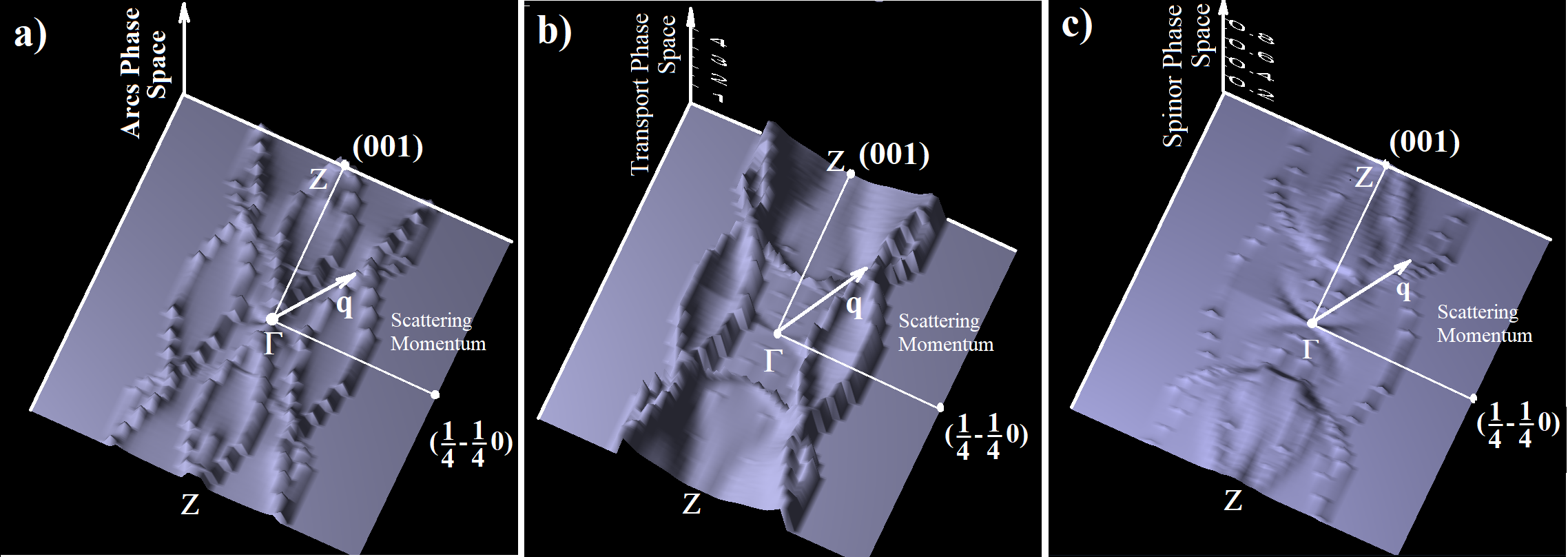}
\caption{Phase space calculation of scattering processes involving Fermi arc electrons in 
Cd$_{3}$As$_{2}$. Case (a) is the double-delta integral, (b) includes the velocity prefactor and (c)
includes both velocity and spin prefactors. Note that the normalization
factor in (a) differs from the other two plots so only (b) and (c) can be
compared quantitatively.}
\label{hiq}
\end{figure*}

We now proceed to the analysis of the surface--mediated transport in Cd$_{3}$%
As$_{2}$ from first principles. The relaxation rate of the electrons due to
impurities or phonons is related to the matrix element of their scattering
potential. This usually results in a temperature--independent impurity
contribution to the resistivity of a metal often seen at very low
temperatures, with an electron--phonon contribution that scales linearly
with temperature, $\rho _{e-ph}(T)\propto \lambda _{tr}T$, where $\lambda
_{tr}$ is the so called transport coupling constant, which
captures the scattering processes of the electrons near the Fermi surface.
It can be obtained from an integral over the Brillouin Zone $\lambda
_{tr}=\sum_{\bm{q}}=\lambda _{tr}(\bm{q})$ \cite{Allen1978}, of
various scattering events subjected to the momentum and energy conservation
laws: 
\begin{equation}
\lambda _{tr}(\bm{q}) \sim \sum_{\bm{k}}(v_{\bm{k}\alpha
}-v_{\bm{k+q}\alpha })^{2}|V_{\bm{kk+q}}|^{2}\delta ({\epsilon _{%
\bm{k}})}\delta ({\epsilon _{\bm{k+q}})}.  \label{EQ1}
\end{equation}
Here, $\epsilon_{\bm{k}}$, $\epsilon_{\bm{k+q}}$ are the energies
and $v_{\bm{k}\alpha }$, $v_{\bm{k+q}\alpha }$ are the Fermi
velocities at points $\bm{k},\bm{k+q}$ in the phase space. The delta
functions $\delta ({\epsilon _{\bm{k}})}$, $\delta ({\epsilon_{\bm{k+q}})}$ 
constrain scattering to the Fermi surface. The transport direction
is set by $\alpha =x,y,z$, while $V_{\bm{kk+q}}$ is the matrix element
of the potential arising from either impurities or atomic displacements
induced by a phonon with wavevector $\bm{q}$. For the spinor states of
the electrons, it has both spin and spatial contributions which can be
accounted for in a factored form for non--magnetic impurities or phonons:
$V_{\bm{kk+q}}=V_{\bm{kk+q}}^{spin}V_{\bm{kk+q}}^{space}.$

Generally, electron--phonon resistivity calculations can be carried out
for real materials from completely first principles \cite{Savrasov_elph,
Resta2018}. The same is true for supercell simulations with impurities,
however, the large unit cell of Cd$_{3}$As$_{2}$ expanded to a 40-layer slab
contains thousands of atoms and would make such direct approaches
computationally intractable. Instead, we use our previously derived
tight-binding fit to perform the analysis of the phase space available to
scattering using Eq.(\ref{EQ1}), and derive our conclusions based on this phase
space calculation.

For a slab extended in the $\langle 110\rangle $ direction, the BZ will be
highly compressed along $k_{\langle 110\rangle }$, therefore we use a 
two-dimensional $800\times 800$ $\bm{k}$-point grid for the integration in Eq.(%
\ref{EQ1}). Since the Fermi arcs are very narrow along the $k_{\langle
1-10\rangle }$ direction, the $\bm{q}$-point region for visualizing $%
\lambda _{tr}(\bm{q})$ is selected to be a part of the BZ, spanned by the
corners $(1/4,-1/4,0)$, $(-1/4,1/4,0)$ in the $(2\pi /a,2\pi /a,2\pi /c)$
units of the reciprocal space.

Before discussing the numerical results, we qualitatively consider the
behavior of the phase space available to scattering. The main
contributions to $\lambda _{tr}(\bm{q})$ are known to be the
back--scattering processes since in this case the electronic velocities entering Eq.(\ref%
{EQ1}) will point in opposite directions. The back--scattering 
($\bm{k}\rightarrow -\bm{k}$) occurs in every three dimensional Fermi 
surface but would be absent for a true Weyl semimetal since its Fermi arcs 
reside on different surfaces. However, this is no longer true in a Dirac 
semimetal where both Fermi arcs appear at the same surface. We however, notice here
that aside from small regions near the Dirac point projections, the spinor
states on opposite Fermi arcs are anti-aligned as seen in Fig. \ref{dirac-combined}b.
This makes such back--scattering processes strongly cancel each other
for non magnetic impurities or phonons, due to the orthogonality of spinors with
opposite spins. In contrast, side-scattering 
($\bm{k}\rightarrow \bm{k^{\prime }}$) 
can occur between any two momentum points along each Fermi
arc. The magnitude of this contribution depends on the Fermi velocity term 
$(v_{\bm{k}\alpha }-v_{\bm{k+q}\alpha })^{2}$, where the velocities
are oriented perpendicular to the arc. For the long, straight Fermi arcs
that we find in Cd$_{3}$As$_{2}$, electrons are only scattered between
states with parallel velocities, thus reducing this contribution to zero. This is analogous to the case of Fermi arc states in Weyl semimetals such as TaAs 
\cite{Resta2018} and NbAs \cite{NbAs2019}.

These contributions can be visualized (Figure \ref{hiq}) in the $\bm{q}$%
-dependence of the phase space integral by adding each term in the
expression (\ref{EQ1}) iteratively. First, we evaluate the bare contribution 
$\sum_{\bm{k}}\delta ({\epsilon _{\bm{k}})}\delta ({\epsilon _{%
\bm{k+q}})}$, shown in Figure \ref{hiq}(a). The effect of the
dumbbell-like structure of the Fermi arcs is apparent here, forming two
bright bands flanking a bowtie-shape along the central strip of the $%
\bm{q}$-space. While this term lacks the velocity pre--factor of the full
expression and cannot be quantitatively compared to the other calculations,
it clearly shows all of the allowed transitions within the phase
space.

Next we add the velocity contribution, computing the integral $\sum_{\bm{%
k}}(v_{\bm{k}\alpha }-v_{\bm{k+q}\alpha })^{2}|\delta ({\epsilon_{%
\bm{k}})}\delta ({\epsilon_{\bm{k+q}})}$, shown in Fig. \ref{hiq}%
(b). This greatly reduces the amplitude of side scattering processes along the Fermi arcs extended in the $q_{z}$ direction,
evidenced by the disappearance of the central bowtie structure in the phase
space. 

We can also quantitatively compare the effect of including the spin term $%
\sum_{\bm{k}}(v_{\bm{k}\alpha }-v_{\bm{k+q}\alpha })^{2}|V_{%
\bm{kk+q}}^{spin}|^{2}\delta ({\epsilon _{\bm{k}})}\delta ({\epsilon
_{\bm{k+q}})}$, since for non--magnetic scatterering processes this reduces to
evaluating the overlap between two spinor states of the electrons. The result
is shown in \ref{hiq}(c), where we also indicate the scales of the obtained
phase space functions since the integrals with and without $|V_{\bm{kk+q}%
}^{spin}|^{2}$ have the same units. As one can see, the effect of including the
spinor overlaps results in the complete suppression of the back--scattering terms,
making their contributions to $\lambda _{tr}$ almost negligible throughout the
entire phase space,  as evidenced in Fig. \ref{hiq}(c) and its comparison
with \ref{hiq}(b) (note the difference in scales).  The only remaining
contributions are faint regions at the edge, which correspond to weak
scattering terms between $\bm{k}$-points located near opposite Dirac
node projections.

The results obtained here can naturally explain recent low temperature
resistivity measurements of Cd$_{3}$As$_{2}$ \cite{Liang2015}, where
it was found that some samples below 5K exhibited very long transport lifetimes, up to 10$^{4}$ longer than the quantum lifetimes. The resistivity
anisotropy was observed to be 20--30 in samples with large lifetime enhancements,
along with measured ultrahigh carrier mobilities that were claimed
to be protected by an "unknown mechanism" \cite{Liang2015}. Although such a
behavior would indeed be unexpected for bulk Dirac cone states in Cd$_{3}$As%
$_{2},$ it is easily understood if the surface transport at very low
temperatures is taken into account. Here, in the absence of thermally
activated carriers at the bulk, the Fermi arc surface states would be the main
contributor to the conductivity, so strong anistoropy and very
large carrier mobility are expected as seen from our phase space calculation
with the strongly suppressed back-- and side--scattering events. 


\section{Topological Insulator Phase}

We now proceed to the analysis of the Cd$_{3}$As$_{2}$ surface states in the
presence of an inversion breaking term. This can occur either in the
vicinity of the contact interface with the electrical leads or when a strong electric field is applied.
Both effects can naturally occur along the  $\langle 110\rangle $ growth
direction, breaking the inversion and $C_{4}$ rotation symmetries of the
system, and introducing a small gap at the Dirac points. This
symmetry--breaking effect can be modeled at the level of the crystal
structure by introducing small opposite shifts of the Cd cations and As
anions along the $\langle 110\rangle $ direction. It
can be easily analyzed based on a $4 \times 4 \bm{k}\cdot \bm{p}$ model Hamiltonian. (See Supplemental Material for details)

\begin{figure}[tbp]
\includegraphics[width=1.0\columnwidth]{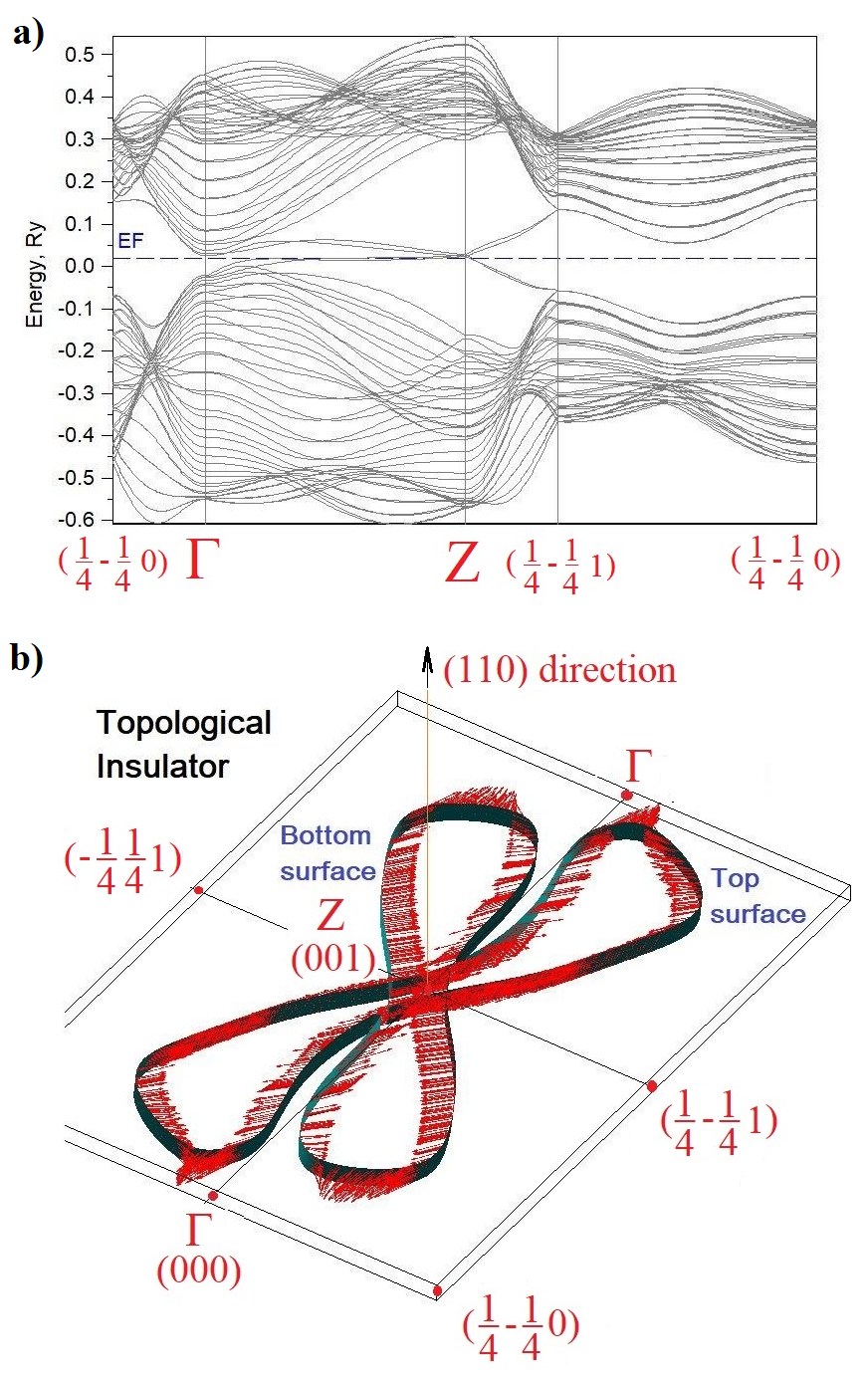}
\caption{40-layer slab calculation of the tight-binding
model fit for the topological insulator phase of Cd$_{3}$As$_{2}$, showing the a) band structure, and b) surface states. Note that the shown region is stretched along the $k_{\langle 1-10\rangle }$ direction in order to more clearly show the surface states.}
\label{ti-combined}
\end{figure}


Gapping a Dirac point along a high symmetry line can result in either a Weyl
semimetal or topological insulator (TI) \cite{Yan2017_review}. We eliminate
the first possibility by employing the monopole mining method \cite{mmm},
which finds no sources/sinks of Berry curvature within the BZ. For the
second case, we determine the classification \cite{Fu2007, fukane} of the TI
phase by explicitly computing the topological indices on each torus using a
discretized plaquette link method \cite{fukui, link-fplapw}. This identifies
a weak TI phase, which is confirmed by our calculation of (a) the surface
energy band states (Figure \ref{ti-combined}a) with the Dirac cone appearing
around zone boundary $Z$ point of the Brillouin Zone, and of the Fermi
surface sheets crossing the edges of the BZ (Figure \ref{ti-combined}b) instead of
encircling the $\Gamma $ point as they would for a strong TI.


The Fermi surface of the Cd$_{3}$As$_{2}$ TI phase is qualitatively very
similar to the Dirac semimetal phase. Fermi surface sheets have long
straight regions extended in the $\Gamma -Z$ direction, as well as the
anti--aligned spin structure at $(\mathbf{k},-\mathbf{k})$ opposite $\mathbf{%
k}$-points. The only deviations from this are slight spin rotations near the 
$\Gamma $ point, where the Dirac node projections were previously located.
Overall however, this Fermi surface structure would similarly suppress both
back and side-scattering processes, resulting in a highly suppressed
scattering just as in the Dirac phase.

\section{Conclusion}

In conclusion, based on our accurate numerical fits to the electronic
structure of Dirac semimetal Cd$_{3}$As$_{2}$, we have computed the surface states in a slab geometry, finding Fermi arcs that stretch through the edges of the Brillouin Zone and produce
very long and straight Fermi surfaces. Their particular shape and spin
structure results in suppression of  both back-- and side-- scattering
effects in the electronic transport, which was explicitly demonstrated by
calculating the available phase space to the scattering for the Fermi arc
electrons of the Dirac semimetal phase. A similar suppression mechanism is
expected for the surface states of a possible weak topological insulator
phase, that can be induced by an inversion--breaking perturbation. Ultra--high
carrier mobility and strong resistivity anisotropy at very low temperatures
naturally emerges from the present study which could explain recent
resistivity measurements in Cd$_{3}$As$_{2}$ \cite{Liang2015}.

Recently, a number of approaches have emerged for engineering topological Dirac states in semiconductor heterostructures \cite{gaas-ge, inas-gasb, insb-sn}. The mature fabrication methods developed around these Group-III, -IV, and -V semiconductors enable the creation of heterostructures with precise control over layer thickness, termination, doping, and strain. Since the shape of Fermi arc surface states is highly dependent on the crystallographic direction of the surface, the atomic terminations, surface strain, and material interfaces \cite{PhysRevB.97.235416, co3sn2s2}, these highly--tunable semiconductor heterostructures could be used to engineer long, straight, Fermi arc states with high mobilities for next--generation electronic devices.

\begin{acknowledgments}
V.I. is supported by startup funding from Virginia Tech.
X.W. was supported by NSFC Grants No. 12188101, 11834006, 12004170, 51721001, and 11790311, as well as the excellent program at Nanjing University. X.W. also acknowledges the support from the Tencent Foundation through the XPLORER PRIZE.
\end{acknowledgments}

\bibliographystyle{plain}

\bibliography{references}

\end{document}